\documentclass{revtex4}
\usepackage{graphicx}
\begin{document}
\title
{Overlapping binary and multiple open cluster candidates in the Galaxy}

\author{G.~Javakhishvili and M.~Todua \
\\Georgian National Astrophysical Observatory, Ilia
State University, Kazbegi ave. 2a, 0160 Tbilisi,
Georgia}

\begin{abstract}
A few binary open clusters with visually separable components are
known in the Galaxy. However there can be dual systems that overlap
in the line of sight and are seen as a single cluster. On the basis
of the WEBDA database we studied the V-band magnitude distributions
of stars of the open clusters in the Galaxy. Most of these
distributions showed Gaussian-like shape with a single peak, but
some of them demonstrated two and more peaks which we considered as
a manifestation of duality or multiplicity of the object. We
examined the magnitude distributions of 563 rich open clusters from
the WEBDA database with more than 100 stars and revealed 70 of them
with two and more statistically significant peaks which we
considered as binary and multiple cluster candidates. On the basis
of these clusters we comprised a catalogue of possible binary and
multiple clusters in the Galaxy overlapping in the line of sight.
Geometric centres of cluster components and corresponding numbers of
cluster members were estimated.
\end{abstract}

%\begin{keywords}(Galaxy:) open clusters  and associations: general,
%catalogues
%\end{keywords}

\maketitle

\section{Introduction}

Studies of binary and multiple clusters are interesting from the
point of view of star formation and evolution in galaxies. A
possible mechanism of formation of double stellar systems had been
suggested by Fujimoto \& Kumai \cite{Fujimoto}, and further
developed by Bekki et al. \cite{Bekki}, where cloud-cloud collisions
were considered as the most feasible way to produce two or more
clusters. A different scenario was proposed by Leon et al.
\cite{Leon} considering a tidal capture.

These theories were introduced after the detection of binary
clusters in the Large and Small Magellanic Clouds (LMC and SMC) in
Bhatia \& Hatzidimitriou \cite{Bhatia1}, Hatzidimitriou \& Bhatia
\cite{Hatzidim}, and references therein. A significant fraction, of
open clusters in the Magellanic Clouds (MCs) were found in pairs,
with the maximum projected separation between the centers of the
components 18.7~pc. The number of found pairs was greater than
expected from projection effects. It was also suggested that some of
the binary cluster candidates were physical pairs. A catalogue with
a photographic atlas of the binary system candidates in the LMC was
compiled by Bhatia et al. \cite{Bhatia2}. Some of these objects
possibly were triple systems and some of the binary cluster
candidates were imbedded in molecular clouds. Hatzidimitriou \&
Bhatia \cite{Hatzidim} catalogued the cluster pairs in the SMC.
Dieball et al. \cite{Dieball} further developed this research and
presented a new catalogue of binary and multiple cluster candidates
in the LMC.

In our Galaxy $\chi$ and $h$ Persei (NGC~869 and NGC~884) had been
for a long time the only distinguishable binary open cluster, until
Subramanyam et al. (1995) detected 18 probable binary systems with
spatial separations between the components of up to 20~pc. Out of
these, 16 pairs had members that seemed to have similar ages, which
might suggest that they were physical pairs. This suggestion was
based on the ages and radial velocities of the member clusters.
These studies revealed that the fraction of clusters in pairs in the
Galaxy was significantly smaller than that in the MC: 9\% from the
400 clusters investigated by Subramaniam et al. (1995), 27\% of the
4089 clusters investigated by Dieball et al. (2002).

All pairs in the above studies were visually separable. However, it
is possible that some cluster pairs overlap in the line of sight and
it is difficult or impossible to resolve them. Thus they are seen as
single clusters. Some of them might be physical pairs, some not.
Galadi-Enriquez et al. \cite{Galadi} considered NGC 1750 and 1758 as
overlapping open clusters, although they suggested that this pair
did not make up a gravitationally bound system.

Studying open clusters from the WEBDA database for membership
determination using the accumulation method, described in
Javakhishvili et al. \cite{Javakh}, we examined the distributions of $V$-band
magnitudes of cluster members. Most of these
distributions displayed Gaussian-like asymmetric shape with a single
peak. However some of them revealed two and more distinct peaks, at
which we decided to take a closer look. We propose that this feature
of magnitude distributions could indicate duality or multiplicity of
the open cluster systems overlapping in the line of sight.

\section{The method}

\begin{figure}
  \centering
  \includegraphics[width= 0.4\textwidth] {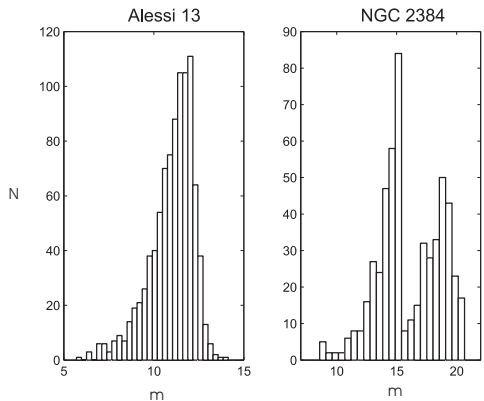}
  \caption{V-magnitude distributions of clusters Alessi~13 and NGC~2384.}
  \label{samples}
\end{figure}

There are 938 open clusters in the WEBDA database with available
rectangular positions (hpd type) of the member stars. Among them we
choose 563 rich clusters with more than 100 members to examine the
distributions of V-magnitude of stars in each cluster. Limiting
magnitudes of stars varied for different clusters from $m_V=13$ to
$25$. Most of the V-magnitude distributions showed a typical
Gaussian-like form with a single peak. However the V-magnitude
distributions of about 70 clusters displayed two and more "humps".
Fig.~\ref{samples} illustrates two examples of these distributions.
The first one - Alessi~13 - has a single magnitude peak, which is
typical for 88\% of the chosen rich clusters. The second one -
NGC~2384 - exhibits two distinct peaks. We propose that one of the
explanations for this kind of profile can be two clusters
overlapping due to projection effect. If this is the case, then they
might be either close physical systems or separate clusters
projected along the same line of sight. Since the probability of the
latter is less than 1 (Subramanyam et al. \% \cite{Sub}), it is
likely that most of them are physically associated.

\begin{figure}
  \centering
  \includegraphics[width=0.4\textwidth] {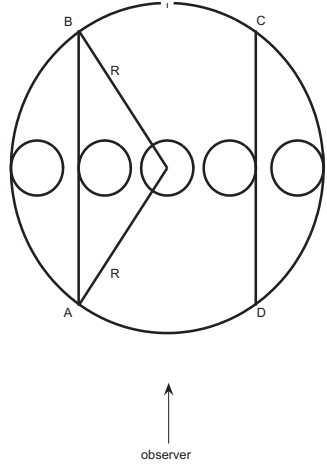}
  \caption{For estimation of the probability to see an overlapping
           binary cluster. The radius of the big circle $R$=20pc, the diameters of small circles are 3pc.}
  \label{project}
\end{figure}

Binary open clusters projected on the sky can be seen with separated
members like h and $\chi$ Persei, but some of them should overlap,
one being behind another, so that the components cannot be resolved
(Fig.~\ref{project} within the volume ABCD). Assuming the maximum
distance between the centres of the two clusters to be 20~pc and
equal sizes of the constituents, we estimated the probability that
an arbitrary pair is seen as an overlapping cluster:

\begin{equation}
%\label{}
  p=\frac{V_{ABCD}}{V_{sphere}}100\%
\end{equation}

\begin{figure}
  \centering
  \includegraphics[width=0.4\textwidth] {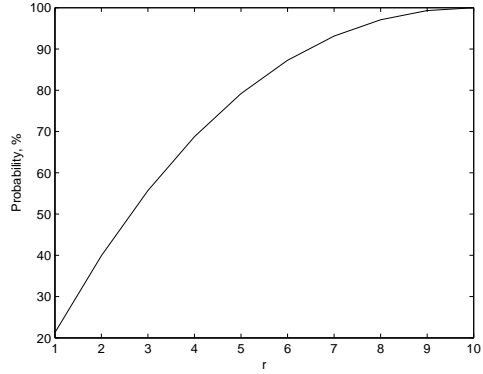}
  \caption{Fraction of the overlapping binary clusters among all binary
           cluster candidates depending on
  the radii ($r$) of the components.}
  \label{probability}
\end{figure}

This probability depends on the sizes of the components and is
plotted in Fig.~\ref{probability}. According to this plot, for
example, about 70\% of the binary clusters, each of the members of
which have diameters about 3~pc should be seen to an observer as
overlapped and thus as a single cluster, and all the clusters with
diameters $>10$~pc should overlap.

\begin{figure}
  \centering
  \includegraphics[width=0.4\textwidth] {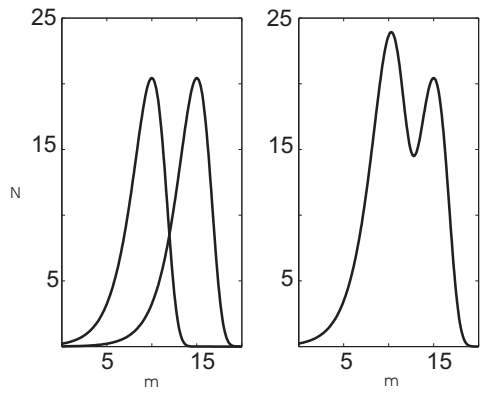}
  \caption{Extreme value distribution functions: two single
           distributions (left) and their superposition (right).}
  \label{ev1ev2}
\end{figure}

The characteristic asymmetric profile of the V-magnitude
distribution of an open cluster, like Alessi~13, can be well fitted
by the extreme value distribution (EVD) function. The profile
similar to that of NGC~2384 with two (or more) peaks can be best
adjusted by the combination of two (or more) EVD functions:

\begin{equation}
\label{evd}
  f(m)=\sum{\frac{A_{i}}{{\sigma_{i}}}e^{\frac{m-\mu_{i}}
  {\sigma_{i}}}e^{-e^{\frac{m-\mu_{i}}{\sigma_{i}}}}}
\end{equation}
$i=1,2,...n$, where $n$ is the number of components, $\mu_{i}$ is a
centre of EVD for i-th component, $A_{i}$ is the fitted coefficient,
and $\sigma_{i}$ the standard deviation. Two EVD functions with
different centres and their superposition are shown in
Fig.~\ref{ev1ev2}.

\section{Discussion}

\subsection{General discussion}

All 70 clusters with two and more "humps" in their V-magnitude
distributions were processed and fitted by the combination of EVD
functions (eq. \ref{evd}). The goodness of fits were in the range
of 0.8 to 0.98. These clusters were compiled into a catalogue
presented in Tables 1 and 2. With our criteria, we found 56
binary, 12 triple and 2 quadruple overlapping cluster candidates.

The number of stars in the i-th cluster member was determined as
\begin{equation}
\label{totnum}
   N_{i}=\sum{f_{i}(m_j)}
\end{equation}
where $f_{i}$ is the EVD function for the i-th cluster, $j$ runs the
bin numbers of the i-th cluster.

Discrepancies between the total number of stars $N$ and the sum of
the calculated numbers of stars in the members were
\begin{equation}
\label{delnum}
  \Delta{N}{\%}=|\frac{N-\sum{N_{i}}}{N}*100|
\end{equation}
which characterize the uncertainty of belonging of some stars to
either components.

Geometrical centres of the clusters and those of the i-th companion
were determined as
\begin{equation}
\label{geo}
   X_{0}=\frac{\sum{x_j}}{N}, \hspace{5mm} Y_{0}=\frac{\sum{y_j}}{N},
    \hspace{5mm} j=1,...,N
\end{equation}

\begin{equation}
\label{geocomp}
   X_{i}=\frac{\sum{x_j}{p_{i}(m)}}{N_i}, \hspace{5mm}
   Y_{i}=\frac{\sum{y_j}{p_{i}(m)}}{N_i}
\end{equation}
where $p_i$ is the probability for a star with magnitude $m$ to
belong to the i-th component:

\begin{equation}
\label{prob}
 p_i(m)=\frac{f_i(m)}{\sum{f_k(m)}}, \hspace{5mm} \hspace{5mm} i,k=1,...,n
\end{equation}

\begin{figure}
  \centering
  \includegraphics[width=0.4\textwidth] {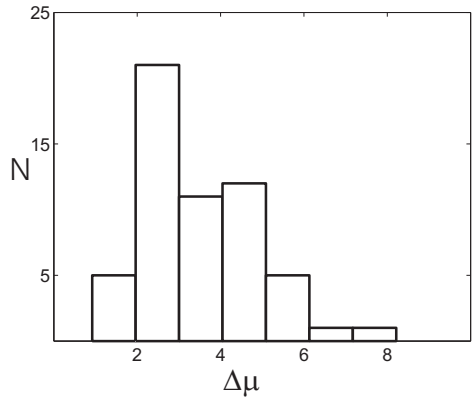}
  \caption{Distribution of differences of the magnitude centres
           $\Delta \mu$ in overlapping binary open cluster candidates.}
  \label{deltam}
\end{figure}

The differences in magnitude peaks $\Delta \mu$ range from 1.5 to 8,
the majority of them falling between 1.5 and 3 (see
Fig.~\ref{deltam}). Differences in intensities between the two
structures are

\begin{equation}
\label{deli}
\centering
  \Delta I = 2.512^{\Delta \mu}
\end{equation}

For $\Delta \mu = 4$, as in the case of NGC~2384 in
Fig.~\ref{samples} (which is a young cluster at a distance of about
2~kpc) $\Delta I \approx 40$ which means that the distance modulus
of the components differ by about 6 times. If we assume that there
is an absorption cloud between the two structures, then their
separation will be considerably smaller. Young clusters are located
in large molecular clouds (Efremov \cite{Efremov}). Some binary
cluster candidates in the catalogue of Bhatia et al. \cite{Bhatia2}
were imbedded in clouds. In this case $\Delta \mu$ could be due to
the inter-cluster cloud absorption. When a cluster formation starts
in two parts of a giant molecular cloud, a light pressure from young
stars drives out the rest of the cloud. These fragments could be
kept by the gravity of the clusters and concentrate in the libration
zones, thus forming the dense absorptive clouds, which could be
responsible for $\Delta \mu$. Though in some cases, if there is
another cluster nearby, the second peak can appear due to stars of
this cluster. This might be true for the magnitude distribution of
NGC~2384, shown on the Fig.~\ref{samples}, which has a nearby
cluster NGC~2383. As was stated in Piskunov et al. \cite{Piskunov},
the central part of NGC~2384 might be relatively free from
contamination from NGC~2383, it is possible that the farer parts of
it might not.

\begin{figure}
  \centering
  \includegraphics[width=0.4\textwidth] {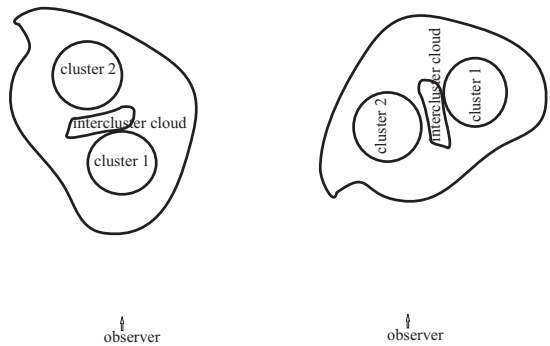}
  \caption{Two possible orientations of a binary open cluster imbedded
            in a molecular cloud (outer curve). The binary cluster members are
            represented by circles, separated by an intercluster cloud.
            These configurations produce different magnitude distributions.}
  \label{configuration}
\end{figure}

Various configurations of projections of clusters and inter-cluster
clouds (Fig.~\ref{configuration}) result in different magnitude
profiles. In the case of the configuration when one companion is
behind another separated by a molecular cloud (on the left diagram in
Fig.~\ref{configuration}), an observer sees a single cluster instead
of two, but the magnitude distribution shows two peaks. In the case
of the configuration on the right in Fig.~\ref{configuration} one
sees a visually separable double cluster.

\begin{figure}
  \centering
  \includegraphics[width=0.4\textwidth] {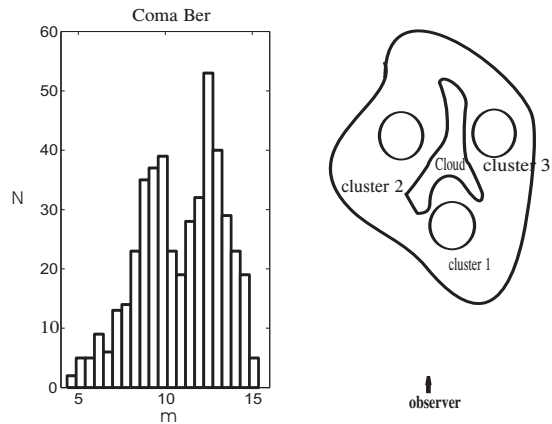}
  \caption{V-magnitude distribution of stars in Coma Ber open cluster
   (left) and one of the possible configurations
   accounting for this histogram (right).}
  \label{tri_cluster1}
\end{figure}

\begin{figure}
  \centering
  \includegraphics[width=0.4\textwidth] {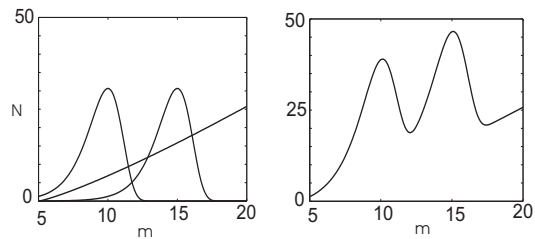}
  \caption{Two components of a binary cluster plus background stars
           (left) and their combination (right).}
  \label{two_background}
\end{figure}

We encountered two cases of the double-peak magnitude distributions:
when the first peak is higher than the second, and vice versa. If
the second peak is higher than the first (the case of Coma Berenices
cluster, Fig.~\ref{tri_cluster1}), it can indicate that either (a)
the rare cluster is more populous than the front one, or (b) there
are two or more clusters in the rare (right picture on
Fig.~\ref{tri_cluster1}), or (c) the tail of the rare cluster is
"contaminated" by a large number of background stars, and therefore
it looks richer (Fig.~\ref{two_background}).

\begin{figure}
  \centering
  \includegraphics[width=0.4\textwidth] {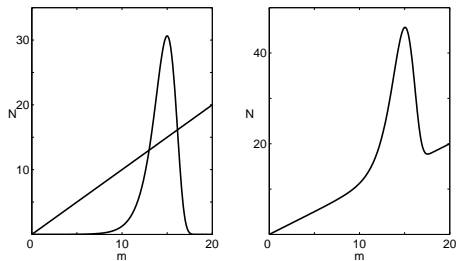}
  \caption{A possible configuration for the magnitude distribution of
           NGC~6633: the second peak is due to
           "contamination" by background stars.}
  \label{clus_back}
\end{figure}

\begin{figure}
  \centering
  \includegraphics[width=0.4\textwidth] {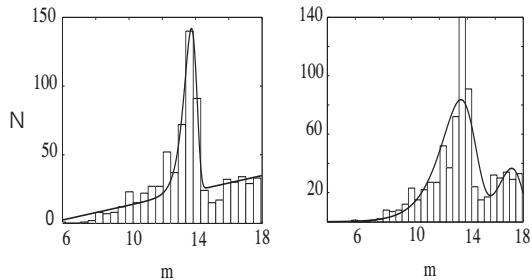}
  \caption{The magnitude distribution of stars in NGC~6633 with two
           different fittings: "cluster + background stars" (left) and
           "cluster + cluster" (right).}
  \label{ngc6633_2}
\end{figure}

We would like to point out the case of NGC~6633.
Fig.~\ref{clus_back} shows how the combination of "cluster +
background stars" can result in the magnitude distribution like that
of NGC~6633 (Fig.~\ref{ngc6633_2}). For the real distribution we
used two different fittings shown in Fig.~\ref{ngc6633_2}: one with
the "cluster + background stars" combination (left diagram), and the
usual combination of two EVDs (right diagram). In both cases the
goodness of fits were high, though it seems that the second peak is
due to the background stars.

Selection effects and incompleteness of data also can affect
magnitude distributions. To determine the probable cause of the
double peak in a magnitude histogram - either duality, or
neighbouring cluster, or background/foreground stars, each case must
be studied individually.

The differences of magnitude centres of the clusters from our list
vary from $\Delta \mu=1$ to 8. Fig.~\ref{deltam} demonstrates the
distribution of this value, where $\Delta \mu \approx 3$ is
dominant.

\begin{figure}
  \centering
  \includegraphics[width=0.4\textwidth] {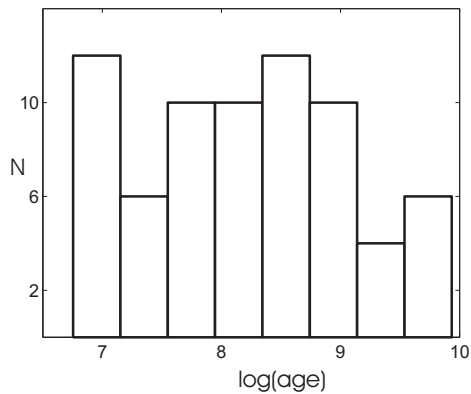}
  \caption{Age distribution of overlapping double and multiple open
           cluster candidates.}
  \label{Age}
\end{figure}

In Fig.~\ref{Age} the distribution of ages of overlapping double and
multiple open clusters in our list are presented, demonstrating that
clusters of all ages are presented almost evenly, about 50\% being
young objects.

\subsection{The case of Coma Berenices cluster}

Studying probable duality or multiplicity of open clusters we
considered only V-magnitude distributions for consistency and the
maximum data completeness, since the V-band data were
available for all objects. However many clusters in our list have
measurements in other filters too. We consider here one particular
case with measurements in B-band, usually available for brighter
stars. We present an example of Coma Berenices (or Mellote 111)
cluster as a typical case of an overlapping double open cluster
candidate in more detail. The distance to the object is 96~pc - this
is the second closest cluster in our list. It is a young cluster
($log(age)=8.7$) and 459 stars are listed in the WEBDA database.

\vspace{5mm}

\begin{figure}
  \centering
  \includegraphics[width=0.4\textwidth] {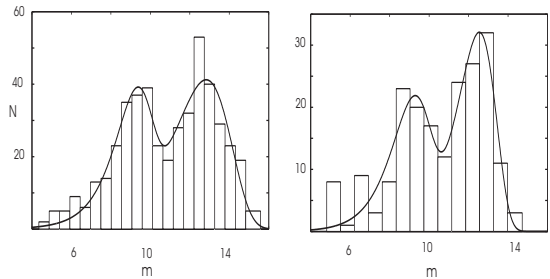}
  \caption{V-magnitude distributions of stars in Coma Ber open
           cluster and its EVD fits: for all stars listed in WEBDA
           (left) and for the stars after membership determination (right).}
  \label{coma_ber}
\end{figure}

The V-magnitude distribution of Coma Berenices cluster with its EVD fit is
demonstrated in Fig.~\ref{coma_ber} (left diagram). The peaks appear
at $\mu=9.2$ and 12.5. The difference between the magnitudes $\Delta
\mu=3.3$ is typical for most of the members of our list. The
goodness of fit is 0.87. To exclude background stars we used the
accumulation method for determination of cluster membership by
Javakhishvili et al. \cite{Javakh}, selecting the stars with close
proper motions. Thus the number of cluster members (having both
proper motions and $B-V$), with probability more than 90\%, were
198. The corresponding V-magnitude distribution
(Fig.~\ref{coma_ber}, right diagram) also reveals dual feature of
the cluster around almost the same magnitude peaks $\mu=9.3$ and
12.7. This indicates that stars around these peaks have similar
proper motions, thus they possibly are physically associated.

\begin{figure}
  \centering
  \includegraphics[width=0.4\textwidth] {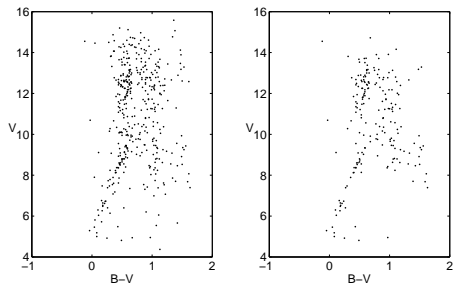}
  \caption{Color-magnitude diagrams of Coma Ber open cluster:
           for all stars listed in WEBDA (left) and for the stars
           after membership determination (right).}
  \label{coma_ber_cmd_all}
\end{figure}

In Fig~\ref{coma_ber_cmd_all} the color-magnitude diagrams (CMD) of
the cluster are presented, for all stars (right diagram) and cluster
members after membership determination (left diagram). Double
feature also can be seen in these CMDs and around the same
V-magnitudes $\mu=9$ and 13, which can be one more argument in
favour of duality of Coma Ber open cluster, the components of which
are visually overlapped.

\vspace{5mm}

\section{Summary}

In addition to the visually separable binary open cluster candidates in the
Galaxy found by Subramaniam et al. (1995), there might be pairs that
overlap in the line of sight and are seen as single clusters. On the
basis of the WEBDA database we investigated open clusters in the
Galaxy considering the distributions of the V-magnitudes of cluster
members. Those clusters that demonstrated two or more magnitude
peaks were regarded as candidates for overlapping binary and
multiple systems. We found that these distributions can be well
fitted by combinations of two or more extreme value distribution
functions. We examined the V-magnitude distributions of 563 clusters
from the WEBDA for which more than 100 members are listed and revealed that 70
of them have more than one peaks and considered them
as binary and multiple cluster candidates. On the basis of these
assumptions we compiled a catalogue of the overlapping binary and
multiple cluster candidates of the Galaxy. Geometric centres of the
component clusters and corresponding numbers of stars were
estimated.

\vspace{5mm}

\textit{Acknowledgement}. In this paper we made use of the WEBDA
database. We would like also to thank Dr.~G.~Didebulidze for useful
comments and suggestions.

\clearpage

\begin{table}
\caption{List of the overlapping binary and multiple cluster
candidates in the Galaxy. General information: cluster identifier,
right ascension, declination, Galactic latitude and longitude,
angular diameter, distance and logarithm of age.}
 \tiny{
\begin{tabular}{|l|c|c|r|r|r|r|r|}
\hline
 cluster  & $\alpha(2000)$ & $\delta(2000)$ & l & b & diameter & distance & log(age) \\
identifier &  h m s & $\circ$  $\prime$ $\prime\prime$ & $\circ$ & $\circ$ & $\prime$  & pc & \\
 \hline
 \multicolumn {8}{|c|}{\normalsize {Binary clusters}} \\
 \hline
        NGC 129  & 00 30 00 & +60 13 06 & 120.270 & -2.543 & 19.0 & 1625 & 7.9 \\
        NGC 188  & 00 47 28 & +85 15 18 & 122.843 & 22.384 & 17.0 & 2047 & 9.6 \\
        King 2   & 00 51 00 & +58 11 00 & 122.874 & -4.688 &  5.0 & 5750 & 9.8 \\
        NGC 381  & 01 08 19 & +61 35 00 & 124.939 & -1.223 &  6.0 & 1148 & 8.5 \\
        NGC 659  & 01 44 24 & +60 40 24 & 129.375 & -1.534 &  5.0 & 1938 & 7.5 \\
        NGC 869  & 02 19 00 & +57 07 42 & 134.632 & -3.741 & 18.0 & 2079 & 7.1 \\
        NGC 884  & 02 22 18 & +57 08 12 & 135.052 & -3.582 & 18.0 & 2345 & 7.0 \\
        NGC 1039 & 02 42 05 & +42 45 42 & 143.658 &-15.613 & 35.0 &  499 & 8.2 \\
        NGC 1245 & 03 14 42 & +47 14 12 & 146.647 & -8.931 &  9.0 & 2876 & 8.7 \\
        IC 348   & 03 44 30 & +32 17 00 & 160.490 &-17.802 &  7.0 &  385 & 7.6 \\
        NGC 1750 & 05 03 55 & +23 39 30 & 179.178 &-10.695 & 20.0 &  630 & 8.3 \\
        NGC 1817 & 05 12 15 & +16 41 24 & 186.156 &-13.096 & 16.0 & 1972 & 8.6 \\
     Berkeley 17 & 05 20 36 & +30 36 00 & 175.646 & -3.648 &  7.0 & 2700 & 10.1 \\
        NGC 1912 & 05 28 40 & +35 50 54 & 172.250 &  0.695 & 20.0 & 1066 & 8.5 \\
        NGC 1960 & 05 36 18 & +34 08 24 & 174.535 &  1.072 & 10.0 & 1318 & 7.5 \\
        NGC 2099 & 05 52 18 & +32 33 12 & 177.635 &  3.091 & 14.0 & 1383 & 8.5 \\
        NGC 2129 & 06 00 41 & +23 19 06 & 186.555 &  0.056 &  5.0 & 1515 & 7.3 \\
        NGC 2158 & 06 07 25 & +24 05 48 & 186.634 &  1.781 &  5.0 & 5071 & 9.0 \\
     Berkeley 73 & 06 22 00 & -06 21 00 & 215.278 & -9.424 &  2.0 &   15 & 9.4 \\
   Collinder 110 & 06 38 24 & +02 01 00 & 209.649 & -1.978 & 18.0 & 1950 & 9.2 \\
      Dolidze 25 & 06 45 06 & +00 18 00 & 211.942 & -1.273 & 20.0 & 6304 & 6.8 \\
        NGC 2301 & 06 51 45 & +00 27 36 & 212.558 &  0.279 & 14.0 &  872 & 8.2 \\
      Tombaugh 1 & 07 00 29 & -20 34 00 & 232.334 & -7.314 &  5.0 & 3000 & 9.0 \\
        NGC 2384 & 07 25 10 & -21 01 18 & 235.390 & -2.393 &  5.0 & 2116 & 6.9 \\
     Berkeley 39 & 07 46 42 & -04 36 00 & 223.462 & 10.095 &  7.0 & 4780 & 9.9 \\
        NGC 2506 & 08 00 01 & -10 46 12 & 230.564 &  9.935 & 12.0 & 3460 & 9.0 \\
        NGC 2539 & 08 10 37 & -12 49 06 & 233.705 & 11.112 &  9.0 & 1363 & 8.6 \\
        NGC 2682 & 08 51 18 & +11 48 00 & 215.696 & 31.896 & 25.0 &  908 & 9.4 \\
        NGC 2818 & 09 16 01 & -36 37 30 & 261.980 &  8.584 &  9.0 & 1855 & 8.6 \\
        NGC 3114 & 10 02 36 & -60 07 12 & 283.332 & -3.840 & 35.0 &  911 & 8.1 \\
        NGC 3324 & 10 37 20 & -58 38 30 & 286.228 & -0.188 & 12.0 & 2317 & 6.8 \\
   Collinder 228 & 10 42 04 & -59 55 00 & 287.668 & -1.047 & 14.0 & 2201 & 6.8 \\
        NGC 3680 & 11 25 38 & -43 14 36 & 286.764 & 16.919 &  5.0 &  938 & 9.1 \\
        Coma Ber & 12 25 06 & +26 06 00 & 221.353 & 84.025 &120.0 &   96 & 8.7 \\
   Collinder 272 & 13 30 26 & -61 19 00 & 307.595 &  1.202 & 10.0 & 2045 & 7.2 \\
        NGC 5606 & 14 27 47 & -59 37 54 & 314.841 &  0.994 &  3.0 & 1805 & 7.1 \\
        NGC 6025 & 16 03 17 & -60 25 54 & 324.551 & -5.884 & 14.0 &  756 & 7.9 \\
        NGC 6087 & 16 18 50 & -57 56 06 & 327.726 & -5.426 & 14.0 &  891 & 8.0 \\
        NGC 6204 & 16 46 09 & -47 01 00 & 338.560 & -1.040 &  5.0 & 1085 & 7.6 \\
        NGC 6253 & 16 59 05 & -52 42 30 & 335.460 & -6.251 &  4.0 & 1510 & 9.7 \\
    Ruprecht 130 & 17 47 32 & -30 06 00 & 359.221 & -0.960 &  3.0 & 2100 & 7.7 \\
        NGC 6451 & 17 50 41 & -30 12 36 & 359.478 & -1.601 &  7.0 & 2080 & 8.1 \\
        NGC 6494 & 17 57 04 & -18 59 06 &   9.894 &  2.834 & 29.0 &  628 & 8.5 \\
        NGC 6633 & 18 27 15 & +06 30 30 &  36.011 &  8.328 & 20.0 &  376 & 8.6 \\
        NGC 6649 & 18 33 27 & -10 24 12 &  21.635 & -0.785 &  5.0 & 1369 & 7.6 \\
         Basel 1 & 18 48 12 & -05 51 00 &  27.355 & -1.947 &  5.0 & 2178 & 7.9 \\
        NGC 6811 & 19 37 17 & +46 23 18 &  79.210 & 12.015 & 14.0 & 1215 & 8.8 \\
         IC 4996 & 20 16 30 & +37 38 00 &  75.353 &  1.306 &  6.0 & 1732 & 6.9 \\
     Berkeley 86 & 20 20 24 & +38 42 00 &  76.667 &  1.272 &  6.0 & 1112 & 7.1 \\
        NGC 6910 & 20 23 12 & +40 46 42 &  78.683 &  2.013 & 10.0 & 1139 & 7.1 \\
        NGC 6939 & 20 31 30 & +60 39 42 &  95.903 & 12.304 & 10.0 & 1185 & 9.3 \\
        NGC 6940 & 20 34 26 & +28 17 00 &  69.860 & -7.147 & 25.0 &  770 & 8.9 \\
         IC 5146 & 21 53 24 & +47 16 00 &  94.383 & -5.495 & 20.0 &  852 & 8.0 \\
        NGC 7243 & 22 15 08 & +49 53 54 &  98.857 & -5.524 & 29.0 &  808 & 8.1 \\
        NGC 7510 & 23 11 03 & +60 34 12 & 110.903 &  0.064 &  6.0 & 2075 & 7.6 \\
    Markarian 50 & 23 15 18 & +60 28 00 & 111.350 & -0.225 &  2.0 & 2114 & 7.1 \\
 \hline
   \multicolumn {8}{|c|}{\normalsize {Triple clusters}} \\
  \hline
        NGC 663  & 01 46 09 & +61 14 06 & 129.467 & -0.941 & 14.0 & 1952 & 7.2 \\
    Berkeley 66  & 03 04 18 & +58 46 00 & 139.434 &  0.218 &  4.0 & 5200 & 9.7 \\
        NGC 1976 & 05 35 16 & -05 23 24 & 209.010 &-19.386 & 47.0 &  399 & 7.1 \\
        NGC 2204 & 06 15 33 & -18 39 54 & 226.014 &-16.107 & 10.0 & 2629 & 8.9 \\
       Haffner 6 & 07 20 06 & -13 08 00 & 227.861 &  0.258 &  6.0 & 3054 & 8.8 \\
        NGC 2420 & 07 38 23 & +21 34 24 & 198.107 & 19.634 &  5.0 & 3085 & 9.0 \\
        NGC 2437 & 07 41 46 & -14 48 36 & 231.858 &  4.064 & 20.0 & 1375 & 8.4 \\
        NGC 2571 & 08 18 56 & -29 45 00 & 249.106 &  3.532 &  8.0 & 1342 & 7.5 \\
        Praesepe & 08 40 24 & +19 40 00 & 205.920 & 32.484 & 70.0 &  187 & 8.9 \\
        NGC 5617 & 14 29 44 & -60 42 42 & 314.670 & -0.100 & 10.0 & 1533 & 7.9 \\
        NGC 5662 & 14 35 37 & -56 37 06 & 316.937 &  3.394 & 29.0 &  666 & 8.0 \\
        NGC 7788 & 23 56 45 & +61 23 54 & 116.434 & -0.782 &  4.0 & 2374 & 7.6 \\
\hline
 \multicolumn {8}{|c|}{\normalsize {Quadruple clusters}} \\
  \hline
        NGC 2548 & 08 13 43 & -05 45 00 & 227.873 & 15.393 & 30.0 &  769 & 8.6 \\
        NGC 6709 & 18 51 18 & +10 19 06 &  42.120 &  4.715 & 14.0 & 1075 & 8.2 \\
  \hline
\end{tabular}
}
\end{table}

\clearpage
\begin{table}
\caption{List of the overlapping binary cluster candidates in the
Galaxy. Characteristics: total number of stars (N), number of stars
in the components ($N_{i}$), discrepancy between the total number of
stars and the sum of numbers of stars in the components
($\Delta{N}\%$), geometrical centres of the system ($X_{0}$,
$Y_{0}$) and its components ($X_{i}$, $Y_{i}$), difference of the
magnitude centres ($\Delta \mu$). For triple and quadruple clusters
$N_{3,4}$, $X_{3,4}$, $Y_{3,4}$ and $\Delta \mu$ between 2nd and 3rd
and 3rd and 4th components are given in the second row of the
appropriate cluster.}
 \tiny{
\begin{tabular}{|l|r|r|r|r|r|r|r|r|r|r|r|}
   \hline
cluster&N&$N_{1}$& $N_{2}$ & $\Delta{N}\%$ & $X_{0}$ & $Y_{0}$ & $X_{1}$&$Y_{1}$&$X_{2}$&$Y_{2}$&$\Delta \mu$ \\
identifier & & & & & & & & & & &\\
   \hline
\multicolumn {12}{|c|}{\normalsize {Double clusters}} \\
\hline
        NGC 129  &  2272 &  1227 & 1044 & 0.1 &    48.85 &    20.75 &     6.12 &    60.85 &    42.73 &   -40.10 &     4.8 \\
        NGC 188  & 10584 &  1767 & 8674 & 1.4 &   -88.38 &   180.61 &   -14.52 &    46.40 &   -73.86 &   134.21 &     4.7 \\
        King 2   &  1037 &   644 &  393 & 0.0 &   -23.65 &    -3.78 &   -19.75 &    -1.63 &    -3.90 &    -2.15 &     2.2 \\
        NGC 381  &  2897 &  1283 & 1553 & 2.1 &   -40.56 &    43.25 &   -10.98 &    27.93 &   -29.58 &    15.32 &     2.2 \\
        NGC 659  &   764 &   209 &  556 & 0.1 &   -18.11 &     7.89 &    -4.69 &     4.56 &   -13.42 &     3.33 &     4.8 \\
        NGC 869  &  3754 &  3029 &  549 & 4.7 &    -9.99 &    -2.62 &    -8.49 &    -2.19 &    -1.49 &    -0.43 &     0.9 \\
        NGC 884  &  3293 &  2617 &  521 & 4.7 &    10.93 &    -2.12 &     8.68 &    -1.32 &     2.25 &    -0.81 &     1.0 \\
        NGC 1039 &   993 &   667 &  319 & 0.7 &     0.17 &   -88.56 &   -20.87 &   -67.97 &    21.04 &   -20.60 &     2.1 \\
        NGC 1245 &  1016 &   614 &  353 & 4.8 &    13.11 &     8.92 &     9.58 &     4.40 &     3.53 &     4.52 &     2.5 \\
         IC 348  &  1929 &   430 & 1506 & 0.5 &     4.05 &    20.97 &     1.08 &     3.22 &     2.97 &    17.75 &     4.5 \\
        NGC 1750 &  7416 &  4392 & 2592 & 5.8 &    -0.37 &     0.51 &     1.90 &     0.66 &    -2.27 &    -0.15 &     3.0 \\
        NGC 1817 &  1891 &  1428 &  474 & 0.6 &    20.19 &   -75.36 &     8.82 &   -49.52 &    11.37 &   -25.84 &     3.5 \\
     Berkeley 17 &  5393 &  4336 & 1124 & 1.2 &   -32.42 &   -23.22 &   -22.39 &   -25.77 &   -10.03 &     2.55 &     3.5 \\
        NGC 1912 &  1274 &   304 &  994 & 1.9 &    -3.17 &   -12.09 &    -1.18 &    -4.61 &    -1.99 &    -7.48 &     2.4 \\
        NGC 1960 &  1394 &   719 &  664 & 0.8 &    -0.37 &     1.53 &     2.44 &    -0.00 &    -2.80 &     1.53 &     4.3 \\
        NGC 2099 &  5027 &  1068 & 3891 & 1.4 &     1.43 &    -1.97 &    -0.12 &     0.06 &     1.55 &    -2.02 &     2.4 \\
        NGC 2129 &   863 &   259 &  584 & 2.3 &   -25.67 &     5.22 &    -9.75 &     3.12 &   -15.92 &     2.10 &     5.9 \\
        NGC 2158 &  5359 &  1498 & 3897 & 0.7 &    -0.84 &    12.14 &     5.92 &     5.12 &    -6.76 &     7.02 &     1.2 \\
     Berkeley 73 &   367 &    73 &  292 & 0.5 &  -146.22 &    92.88 &   -15.52 &    23.29 &  -130.70 &    69.59 &     2.2 \\
   Collinder 110 &  2006 &   616 & 1430 & 2.0 &   -60.18 &  -173.05 &   -18.32 &   -29.13 &   -41.86 &  -143.92 &     3.5 \\
      Dolidze 25 &   136 &    29 &  106 & 0.7 &    79.32 &    33.77 &    -7.71 &    60.85 &    87.03 &   -27.08 &     5.8 \\
        NGC 2301 &  1611 &  1002 &  568 & 2.6 &    33.47 &    -1.60 &    17.64 &    -7.53 &    15.83 &     5.94 &     2.7 \\
      Tombaugh 1 &  1970 &  1193 &  740 & 1.9 &   294.07 &   239.12 &   232.01 &   188.79 &    62.06 &    50.33 &     3.0 \\
        NGC 2384 &   549 &   235 &  268 & 8.4 &    -6.49 &     5.47 &    19.87 &     0.97 &   -26.36 &     4.50 &     4.1 \\
     Berkeley 39 &  4408 &  1642 & 2794 & 0.6 &   -45.72 &    15.86 &   -21.24 &    15.83 &   -24.48 &     0.02 &     3.0 \\
        NGC 2506 &  1232 &   217 & 1021 & 0.5 &     2.13 &    -7.26 &     1.31 &    -0.79 &     0.82 &    -6.48 &     2.1 \\
        NGC 2539 &   821 &   356 &  473 & 1.0 &     0.50 &   -34.47 &    -1.94 &   -21.17 &     2.44 &   -13.31 &     4.3 \\
        NGC 2682 &  3284 &  2255 & 1025 & 0.1 &    -0.09 &    -2.49 &    -0.01 &    -2.13 &    -0.08 &    -0.36 &     4.4 \\
        NGC 2818 &  1168 &   188 &  976 & 0.3 &   -11.25 &    -8.19 &    -0.93 &    -2.37 &   -10.32 &    -5.83 &     3.6 \\
        NGC 3114 &  2119 &   230 & 1847 & 2.0 &   -18.41 &    55.30 &    -1.04 &     4.98 &   -17.37 &    50.33 &     8.7 \\
        NGC 3324 &   988 &   626 &  376 & 1.4 &   -24.02 &   138.98 &    -5.99 &    70.31 &   -18.03 &    68.67 &     4.3 \\
   Collinder 228 &  1471 &   328 & 1087 & 3.8 &   -67.11 &    15.21 &     4.77 &    11.51 &   -71.88 &     3.70 &     6.0 \\
        NGC 3680 &   888 &   733 &   70 & 9.6 &    -1.41 &    -2.93 &    -2.02 &    -2.78 &     0.61 &    -0.14 &     1.3 \\
        Coma Ber &   459 &   142 &  305 & 2.6 &   159.96 &   -18.32 &    23.07 &     4.06 &   136.89 &   -22.38 &     3.6 \\
   Collinder 272 &  1826 &   786 &  990 & 2.7 &   129.61 &   -12.88 &    89.05 &   -34.81 &    40.56 &    21.93 &     2.9 \\
        NGC 5606 &   202 &    69 &  130 & 1.5 &    -3.23 &    -8.36 &    -0.14 &    -7.14 &    -3.09 &    -1.22 &     3.9 \\
        NGC 6025 &   182 &   153 &   29 & 0.0 &     0.25 &   -13.13 &    -2.11 &    -9.69 &     2.35 &    -3.44 &     3.0 \\
        NGC 6087 &  1393 &   662 &  686 & 3.2 &     2.37 &    -4.88 &     1.49 &    -4.32 &     0.88 &    -0.55 &     4.0 \\
        NGC 6204 &   523 &   141 &  369 & 2.5 &    83.31 &     9.39 &    39.96 &    11.65 &    43.35 &    -2.26 &     3.6 \\
        NGC 6253 &  8194 &  3207 & 4742 & 3.0 &    68.86 &    43.48 &    73.60 &    31.36 &    -4.74 &    12.12 &     3.5 \\
    Ruprecht 130 &   523 &   310 &  211 & 0.4 &   106.95 &   -70.14 &    86.08 &   -82.54 &    20.86 &    12.40 &     2.5 \\
        NGC 6451 &  1491 &  1128 &  322 & 2.8 &   -15.84 &   -82.21 &   -19.00 &   -78.95 &     3.16 &    -3.26 &     2.1 \\
        NGC 6494 &   332 &   256 &   82 & 1.8 &    -3.27 &    -3.90 &    -0.52 &    -1.76 &    -2.75 &    -2.15 &     2.4 \\
        NGC 6633 &   759 &   259 &  513 & 1.7 &     5.90 &     5.13 &     0.21 &    -0.79 &     5.69 &     5.92 &     2.6 \\
        NGC 6649 &   517 &   125 &  396 & 0.8 &     2.39 &     7.71 &     1.24 &     4.18 &     1.16 &     3.53 &     3.9 \\
        Basel 1  &  1297 &   732 &  550 & 1.2 &     0.14 &    79.64 &    26.99 &    65.48 &   -26.85 &    14.16 &     2.1 \\
        NGC 6811 &   430 &   367 &   52 & 2.6 &    26.35 &    17.61 &    26.62 &    16.08 &    -0.27 &     1.53 &     1.3 \\
        IC 4996  &   775 &   175 &  600 & 0.0 &    43.49 &    51.38 &     9.49 &    -2.29 &    34.00 &    53.67 &     4.7 \\
     Berkeley 86 &   736 &   204 &  518 & 1.9 &    -7.64 &     2.10 &    -8.54 &     2.64 &     0.90 &    -0.54 &     5.9 \\
        NGC 6910 &   401 &   147 &  250 & 1.0 &     3.41 &     2.81 &     0.14 &     0.21 &     3.27 &     2.60 &     3.4 \\
        NGC 6939 &  2870 &   424 & 2495 & 1.7 &   -33.32 &    22.21 &     7.33 &     0.27 &   -40.66 &    21.93 &     6.8 \\
        NGC 6940 &  1054 &   710 &  349 & 0.5 &     0.70 &     0.76 &     0.29 &     0.85 &     0.41 &    -0.09 &     5.1 \\
        IC 5146  &   856 &   659 &  174 & 2.7 &     0.72 &   -12.21 &     4.49 &    -8.76 &    -3.77 &    -3.45 &     3.0 \\
        NGC 7243 &  3194 &   882 & 2329 & 0.5 &    -2.89 &     6.71 &    -0.56 &     0.23 &    -2.33 &     6.48 &     2.0 \\
        NGC 7510 &   839 &   152 &  705 & 2.2 &    -2.25 &     6.01 &    -0.60 &     1.65 &    -1.65 &     4.36 &     2.9 \\
    Markarian 50 &  1262 &   375 &  886 & 0.1 &    18.20 &     9.09 &     6.13 &     0.47 &    12.06 &     8.62 &     4.7 \\
\hline
\end{tabular}
}
\end{table}

\clearpage
\begin{table}
\caption{List of the overlapping triple and quadruple  cluster
candidates in the Galaxy. Characteristics: total number of stars
(N), number of stars in the components ($N_{i}$), discrepancy
between the total number of stars and the sum of numbers of stars in
the components ($\Delta{N}\%$), geometrical centres of the system
($X_{0}$, $Y_{0}$) and its components ($X_{i}$, $Y_{i}$), difference
of the magnitude centres ($\Delta \mu$). For triple and quadruple
clusters $N_{3,4}$, $X_{3,4}$, $Y_{3,4}$ and $\Delta \mu$ between
2nd and 3rd and 3rd and 4th components are given in the second row
of the appropriate cluster.}
 \tiny{
\begin{tabular}{|l|r|r|r|r|r|r|r|r|r|r|r|}
   \hline
cluster&N&$N_{1}$& $N_{2}$ & $\Delta{N}\%$ & $X_{0}$ & $Y_{0}$ & $X_{1}$&$Y_{1}$&$X_{2}$&$Y_{2}$&$\Delta \mu$ \\
identifier & & & & & & & & & & &\\
   \hline

\multicolumn {12}{|c|}{\normalsize {Triple clusters}}\\
    \hline
        NGC 663  & 4560 &  345 & 2282 & 1.3 & -21.74 &  1.74  &  -1.82 & -1.55 &  -8.93 & -2.77 & 3.6 \\
             &       & 1874 &         &      &       &        & -11.00 &  6.06 &        &       & 2.8 \\
     Berkeley 66 & 1677 &  329 &  521 & 0.8 & -67.44 & -91.00 & -76.44 & 28.73 &  -1.69 &-40.56 & 3.0 \\
             &       &  813 &      &     &           &        &  10.68 &-79.17 &        &       & 2.0 \\
        NGC 1976 & 6034 & 1387 & 4357 & 0.1 &-567.48 & -97.37 &-266.45 &-72.67 &-301.21 &-25.53 & 3.5 \\
             &       &  296 &      &      &        &          &   0.18 &  0.84 &        &       & 6.7 \\
        NGC 2204 & 2468 &  649 &  503 & 0.5 &   5.99 &  15.30 &  -1.15 &  2.68 &  -0.02 &  4.57 & 2.3 \\
             &       & 1305 &      &       &        &         &   7.15 &  8.05 &        &       & 1.9 \\
       Haffner 6 &  711 &   77 &   92 & 1.0 &  -9.57 &  21.01 &   0.56 &  2.29 &  -1.18 &  3.29 & 2.3 \\
             &       &  535 &      &      &        &          &  -8.96 & 15.43 &        &       & 2.0 \\
        NGC 2420 &  942 &  204 &  357 & 2.2 &   8.95 &   1.18 &   1.86 & -1.41 &   7.40 &  5.03 & 2.0 \\
             &       &  360 &      &      &        &          &  -0.30 & -2.44 &        &       & 1.9 \\
        NGC 2437 & 1687 &  459 &  409 & 0.1 &  70.18 &  55.90 &   5.98 &  4.54 &  15.93 & 12.93 & 2.9 \\
             &       &  817 &      &       &        &         &  48.27 & 38.43 &        &       & 2.2 \\
        NGC 2571 & 1710 &  127 & 1416 & 1.3 &-128.80 &   2.53 &  -2.97 &  1.49 &-113.70 &  0.74 & 5.4 \\
             &       &  145 &      &      &        &          & -12.14 &  0.31 &        &       & 1.7 \\
        Praesepe & 1689 &  816 &  513 & 2.2 & -21.30 &-115.95 &  -2.75 &-60.44 & -18.56 &-55.51 & 4.3 \\
             &       &  323 &      &      &        &          &   0.00 & -0.00 &        &       & 2.5 \\
        NGC 5617 & 1447 &  381 &  317 & 0.5 &  -9.36 &  -5.81 &  -7.04 & -1.55 &   0.08 & -3.64 & 2.9 \\
             &       &  742 &      &      &        &          &  -2.40 & -0.62 &        &       & 2.1 \\
        NGC 5662 &  978 &  190 &  334 & 1.8 &   0.29 &  -0.54 &  -0.23 & -0.54 &   0.45 &  0.03 & 2.3 \\
             &       &  436 &      &      &        &          &   0.07 & -0.03 &        &       & 4.8 \\
        NGC 7788 &  424 &  120 &   88 & 1.2 & -44.80 & -63.55 &  20.09 &-28.62 & -18.56 & -8.70 & 1.8 \\
             &       &  211 &      &      &       &           & -46.33 &-26.23 &        &       & 1.3 \\
\hline
\multicolumn {12}{|c|}{\normalsize {Quadruple clusters}} \\
\hline
        NGC 2548 & 2247 & 112 &  445 & 1.2 &-9.52 &-9.48 & -0.22 & 1.32 & -3.86 &-1.39 & 2.0   \\
           &         &    516 &  1200 &     &      &      & -3.96 &-4.02 & -1.48 &-5.38 & 1.8  3.3\\
        NGC 6709 & 1921 & 196 &  223 & 0.5 & 1.38 &-1.60 & -0.22 & 0.43 &  1.01 &-0.92 & 1.6   \\
           &         &    216 & 1276 &     &      &      &  0.31 & 0.12 &  0.27 &-1.23 & 0.8  2.9\\
   \hline
\end{tabular}
}
\end{table}


\begin{thebibliography}{}

\bibitem[2004]{Bekki}     Bekki, K., Beasley, M. A., Forbes, D. A., Couch, W. J. 2004, ApJ,
602, 730
\bibitem[1988]{Bhatia1}   Bhatia, R. K., Hatzidimitriou, D. 1988, MNRAS, 230, 215
\bibitem[1991]{Bhatia2}   Bhatia, R. K., Read, M. A., Tritton, S., Hatzidimitriou, D. 1991, A\&AS, 87, 335
\bibitem[2002]{Dieball}   Dieball, A., Muller, H., Grebel, E. K. 2002, A\&A, 391, 547
\bibitem[1986]{Efremov}   Efremov, Y. N. 1986, Astronomicheskij tsirkulyar, 1447
\bibitem[1997]{Fujimoto}  Fujimoto, M., Kumai, Y. 1997, AJ, 113, 249
\bibitem[1998]{Galadi}    Galadi-Enriquez, D., Jordi, C., Trullols, E., and Ribas, I. 1998, A\&A, 333, 471
\bibitem[2006]{Javakh}    Javakhishvili, G., Kukhianidze, V., Todua, M., Inasaridze, R. 2006, A\&A, 447, 3, 915
\bibitem[1990]{Hatzidim}  Hatzidimitriou, D., Bhatia, R. K. 1990, A\&A, 230, 11
\bibitem[1999]{Leon}      Leon, S., Bergond, G., Vallenari, A. 1999, A\&A, 344, 450
\bibitem[2004]{Piskunov}  Piskunov, A. E.; Belikov, A. N.; Kharchenko, N. V.; Sagar, R.; Subramaniam, A. 2004, MNRAS, 349, 1449
\bibitem[1995]{Sub}       Subramaniam, A., Gorti, U., Sagar, R., Bhatt, H.C. 1995 A\&A, 302, 86
\bibitem[2007]{webda}     WEBDA catalogues http://www.univie.ac.at/webda

\end{thebibliography}
\end{document}